# The effect of debt on corporate profitability
# Evidence from French service sector

Mazen KEBEWAR[1]
mazen.kebewar@univ-orleans.fr
mazen.kebewar@gmail.com

Current study aims to provide new empirical evidence on the impact of debt on corporate profitability. This impact can be explained by three essential theories: signaling theory, tax theory and the agency cost theory. Using panel data sample of 2240 French non listed companies of service sector during 1999-2006. By utilizing generalized method of moments (GMM) econometric technique on three measures of profitability ratio (PROF1, PROF2 and ROA), we show that debt ratio has no effect on corporate profitability, regardless of the size of company (VSEs, SMEs or LEs).

---

[1] Laboratoire d'Economie d'Orléans – UMR CNRS 7322 Faculté de Droit, d'Economie et de Gestion, Rue de Blois, B.P. 26739 – 45067 Orléans Cedex 2 – France.

# 1. Introduction

Explaining role of debt in firms' performance is one of the primary objectives of contemporary researches for more than fifty years (Modigliani and Miller 1958). However, this role remains a questionable subject which attracts the attention of many researchers as Goddard et al. (2005), Berger and Bonaccorsi (2006), Rao et al. (2007), Baum et al. (2007), Weill (2008), Nunes et al. (2009), Margaritis and Psillaki (2010) and Kebewar (2012).

Indeed, researchers analyze the debt ratio and try to determine whether an optimal debt ratio exists or not. Optimal debt ratio is generally defined as the one which minimizes the cost of capital for the company, while maximizing the value of company. In other words, the optimal debt ratio is the one which maximizes the profitability of company.

Besides, the divergence between researchers can be observed in theoretical strand of literature. There are three essential theories which highlight the influence of debt on corporate profitability, namely: signaling theory, the agency costs theory and tax theory. First, according to signaling theory, the debt; in the presence of asymmetric information, should be correlated positively to profitability. According to the agency costs theory, there are two contradictory effects of debt on profitability, firstly it is positive in the case of agency costs of equity between shareholders and managers, secondly it's effect is negative, resulting from the agency costs of debt between shareholders and lenders. Finally, the influence of taxation is complex and difficult to predict because it depends on the principles of tax deductibility of interest, income tax and non-debt tax shield.

Furthermore, the disagreement exists not only in the theoretical literature but also it is present in the empirical strand. A negative effect of debt on profitability was confirmed by Majumdar and Chhibber (1999), Eriotis et al. (2002), Ngobo and Capiez (2004), Goddard et al. (2005), Rao et al. (2007), Zeitun and Tian (2007) and Nunes et al. (2009). On the other hand, Baum et al. (2006) & (2007), Berger and Bonaccorsi (2006), Margaritis and Psillaki (2007) & (2010), showed a positive influence. In addition, Simerly and LI (2000), Mesquita and Lara (2003) and Weill (2008), find both effects in their studies. Besides that, Berger and Bonaccorsi (2006), Margaritis and Psillaki (2007) and Kebewar (2012) finds the presence of a non linear effect (inverse U-shaped relationship). Finally, a non significant effect was confirmed by Baum et al. (2007) in American industrial companies.

Several factors may reveal reasons for the contradiction of results in empirical studies. First, these empirical studies focus on different types of sample (countries, sectors, companies and periods). Furthermore, researchers have used different measures of profitability as a dependent variable[1] and various debt ratios as independent variable[2]. Finally, these studies applied different methodologies[3].

The empirical literature concerning the impact of debt on profitability leads us to make two inferences. The first one is that most of the empirical studies focused on listed companies. The second one is related to paucity of studies on the French companies as mentioned by; Goddard et al. (2005), Weill (2008), Margaritis and Psillaki (2010) and recently in Kebewar (2012). These two avenues motivated our study. Moreover, current work is very important because debt is a risky choice whose consequences on the corporate profitability can be considerable (e.g. the risk of bankruptcy and its consequences for the stakeholders). So we will try to find, empirically, the effect of debt on profitability for French not listed companies. In addition, to improve the precision of estimation by reducing the heterogeneousness between sizes of companies, we study the behavior of these firms according to their size. Moreover, we will analyze not only the linear effect of debt on profitability, but also the non-linear effect by estimating a quadratic model which takes into account the squared of variable of debt in the equation of regression.

To do this, we will implement the generalized method of moments (GMM) estimation model on a sample of 2240 firms of service sector observed over the period (1999-2006); these companies are divided into three 'size' classes VSBs (very small business), SMEs (small and medium enterprise) and LEs (large enterprise). According to the proponents of GMM model, it provides solution to the problems of simultaneity bias, reverse causality (especially between profitability and debts) bias and the conundrum of possible omitted variables.

The structure of this paper is as follows. First, we discuss the characteristics of the sample and variables. Then, we present the empirical results. Finally, we wrap up the work with main findings and conclusions.

---

[1] ROA, ROE, ROI, PROF, Tobin's Q, Profit on sales, business performance, VRS: Technical Efficiency, CRS: Technical Efficiency, Profit Margin, Frontier efficiency and BTI: ratio earnings before interest and tax to total assets.
[2] Ratio of total debt, ratio of short-term debt, and ratio of long-term debt.
[3] OLS, GLS, DLS, Weighted least squares, fixed effect, random effect, variance decomposition model, covariance model, maximum likelihood, Method of simultaneous equations, quantile regression and GMM.

## 2. Data

### 2.1 Data description

The sample, which is obtained from the Diane database, consists of an unbalanced panel of 2240 French companies of service sector, over the period of 1999-2006. Our sample is composed of unlisted companies like Limited Companies and Limited Liability Companies. In addition, these companies belong to three classes of size (VSEs, SMEs and LEs)[1].

Public enterprises are excluded from the study because of their special political leverage. Furthermore, we do not include companies with negative equity. In addition, outliers were removed according the procedure of Kremp (1995)[2]. Thus, descriptive statistics and the distribution of firms by size are shown in Table (1)[3].

**Table (1)** DESCRIPTIVE STATISTICS FOR VARIABLES

| CLASS SIZE | CLASS (1) 1 -19 | CLASS (2) 20 - 249 | CLASS (3) 250 - 4999 | TOTAL |
|---|---|---|---|---|
| Number of companies | 1241 | 921 | 78 | 2240 |
| Number of observations | 7001 | 5441 | 451 | 12893 |
| PROF1 | 0,091 (0,088) | 0,078 (0,082) | 0,073 (0,061) | 0,085 (0,086) |
| PROF2 | 0,098 (0,087) | 0,086 (0,080) | 0,078 (0,056) | 0,092 (0,084) |
| ROA | 0,072 (0,073) | 0,059 (0,065) | 0,059 (0,053) | 0,067 (0,070) |
| DT | 0,590 (0,202) | 0,607 (0,192) | 0,619 (0,164) | 0,598 (0,197) |
| TANG | 0,190 (0,195) | 0,227 (0,264) | 0,199 (0,223) | 0,207 (0,229) |
| TAX | 0,176 (0,132) | 0,208 (0,146) | 0,218 (0,156) | 0,191 (0,140) |
| GROWTH | 0,056 (0,221) | 0,070 (0,196) | 0,079 (0,179) | 0,063 (0,209) |

Note: Values without parentheses are the averages and those in parentheses are the standard deviation.

### 2.2 Variables

*2.2.1 Dependent variable*

According to the literature, corporate profitability can be measured by several methods. In the context of our study and to compare our results, we use three measures of profitability:

---

[1] According to the Classification of INSEE: very small enterprises (VSEs) which employs less than 20 employees, small and medium enterprises (SMEs) which employ between 20 and 249 employees, and finally (LEs) whose size is between 250 and 4999 employees.
[2] We deleted the observations which are situated outside the interval defined by the first and third quartiles more or less five once the distance interquartile.
[3] The composition of sample could be provided by requesting the author.

PROF1, PROF2 and ROA. PROF1 is measured by dividing net income from operations by total assets. PROF2 is calculated by dividing earnings before interest and tax to total assets. Return on Assets (ROA) is measured by dividing net income from operations (to which income taxes are subtracted) by total capital.

*2.2.2 Explanatory Variables*

> Debt

In theory, debt ratio can be measured in different ways i.e. total debt ratio, debt ratio as short, medium and long term. In our study, we define the total debt ratio (DT) by dividing the sum of the short and long-term debt by the total assets.

> Tangibility

Tangibility has two conflicting effects on profitability. On the one hand, we expect a positive effect by Himmelberg et al. (1999); they show that tangible assets are easily monitored and provide good collateral and thus they tend to mitigate agency conflicts between shareholders and creditors. On the other hand, we predict a negative correlation, because firms with high levels of tangible assets tend to be less profitable. Firms with high levels of intangible assets (in form of liquidity) have more investment opportunities in the long term, innovation and research and development (Deloof, 2003, and Nucci et al., 2005). The negative relationship between tangibility and profitability has been confirmed in number of studies as Rao et al. (2007), Zeitun and Tian (2007), Weill (2008) and Nunes et al. (2009). In addition, Majumdar and Chhibber (1999) and Margaritis and Psillaki (2007) find a positive relationship. To determine the effect of tangibility on profitability, we use the ratio (TANG); it is calculated by dividing the sum of net tangible assets to total assets.

> Tax

The tax impact on profitability of a company is difficult to predict, because it depends on the principle of tax deductibility of interest on debt. So, if a company does not take advantage of this principle, we expect a negative effect of tax on profitability. On the contrary, if a company takes advantage of this principle, this impact will be positive or not significant. Zeitun and Tian (2007) showed a positive effect of tax on profitability. The impact of tax on corporate profitability is highlighted by using the tax ratio in the regression equation. This ratio (TAX) is calculated by dividing the tax paid to earnings before interest and taxes.

> ➢ Growth opportunities

It is expected that firms having high growth opportunities have a high rate of return, because these companies are able to generate more profits from the investment. Therefore, growth opportunities should positively influence profitability. The positive impact of growth opportunities on profitability is confirmed by most empirical studies such as Psillaki and Margaritis (2007), Zeitun and Tian (2007) and Nunes et al (2009). On the other hand, Margaritis and Psillaki (2010) find a negative effect only in the French chemical sector. Several measures to calculate growth opportunity for companies exists in literature. But in the context of our analysis, we use the ratio of growth opportunity (GROWTH) which is measured by the change in total assets from one year to another.

## 3. Methodology

The model in order to analyze the impact of debt on corporate profitability is as follows:

$$PROF_{i,t} = \beta_0 + \beta_1 DT_{i,t} + \beta_2 TANG_{i,t} + \beta_3 TAX_{i,t} + \beta_4 GROWTH_{i,t} + \sum_{n=1}^{9} \beta_n dumt_n + \eta_i + \varepsilon_{it}$$

Where, subscript 'i' denote the studied company and subscript 't' represent the time period. The dependent variable is the ratio of profitability (PROF1, PROF2 or ROA). Moreover, (DT), (TANG), (TAX) and (GROWTH) represent the ratios of debt, tangibility, tax and growth opportunities. Further, influence of time is taken into account by the introduction of annual dummies (dumt) that capture the specific year effect (1999-2006). The individual fixed effect on companies is represented by the term ($\eta_i$). Finally, the error term which is assumed to be independent and identically distributed (iid) which is represented by the term ($\varepsilon_{it}$).

Regarding the effect of non-linearity between debt and profitability, we estimate a quadratic model which takes into account the debt variable squared in the regression equation. Thus, the model to estimate in this context is as follows:

$$PROF_{i,t} = \beta_0 + \beta_1 DT_{i,t} + \beta_2 DT\^2_{i,t} + \beta_3 TANG_{i,t} + \beta_4 TAX_{i,t} + \beta_5 GROWTH_{i,t} + \sum_{n=1}^{9} \beta_n dumt_n + \eta_i + \varepsilon_{it}$$

The null hypothesis of linearity effect is to test: (H0 : $\beta_2 = 0$). If this hypothesis is rejected, we can conclude the existence of non-linearity between debt and profitability. According to the agency cost theory, the effect of debt on profitability must be positive when ($\beta_1 > 0$ and $\beta_1 + 2\beta_2 DT_{i,t} > 0$). However, if the debt ratio arrives at an adequately high level, this effect can become negative. So, our quadratic specification is consistent with the possibility that the relationship between debt and profitability may not be monotonic, it may switch from positive to negative at a high level of debt. Debt will have a negative impact on profitability when ($DT_{i,t} < -\beta_1/2\beta_2$). A sufficient condition for the inverse U-shaped relationship between debt and profitability to hold is that ($\beta_2 < 0$).

We suspect problems of endogeneity in the estimation equation related to causality of exogenous variables to the dependent variable (especially the debt variable). Therefore, traditional econometric methods such as Ordinary Least Square (OLS), fixed effect and Generalized Least Square (GLS) do not allow us to obtain efficient estimates of such model. So, to solve this problem, we introduce the generalized method of moments on panel (GMM) proposed by Arellano and Bond (1991), Arellano and Bover (1995) and Blundell and Bond (1998). This method can provide solutions to simultaneity bias, reverse causality (especially between debt and profitability) and possible omitted variables. Moreover, it can control the individual and temporal specific effects. Indeed, GMM method is used to solve the problems of endogeneity not only at the debt variable, but also for other explanatory variables by using a series of instrumental variables generated by lagged variables.

The model is estimated by two-step System GMM. In order to choose the best model specification, we examined several specifications according to different assumptions about the endogeneity of variables.

## 4. Results

### 4.1 Descriptive analysis

Table (2) reports the changes in profitability ratios. We note a small decrease in profitability ratio (especially for PROF1 and PROF2). Regarding the evolution of profitability according to the size, we find that very small enterprises (VSEs) realize profitability higher than the small and medium enterprises (SMEs) and the large enterprises (LEs); that is to say, there is inverse relationship over the period between size and profitability. Moreover, we note that the decrease in profitability concerns only to the very small enterprises (VSEs) whereas other enterprises have profitability almost stable over the study period.

**Table (2)** CHANGES IN THE RATIO OF PROFITABILITY

| YEAR | PROF1 | PROF2 | ROA | (PROF1) | | | (PROF2) | | | (ROA) | | |
|---|---|---|---|---|---|---|---|---|---|---|---|---|
| | | | | VSEs | SMEs | LEs | VSEs | SMEs | LEs | VSEs | SMEs | Les |
| 1999 | 0,091 | 0,096 | 0,066 | 0,098 | 0,083 | 0,076 | 0,102 | 0,091 | 0,083 | 0,071 | 0,061 | 0,063 |
| 2000 | 0,087 | 0,093 | 0,065 | 0,096 | 0,077 | 0,079 | 0,101 | 0,085 | 0,080 | 0,071 | 0,058 | 0,064 |
| 2001 | 0,087 | 0,092 | 0,069 | 0,094 | 0,080 | 0,070 | 0,100 | 0,085 | 0,070 | 0,075 | 0,062 | 0,060 |
| 2002 | 0,080 | 0,087 | 0,064 | 0,088 | 0,071 | 0,074 | 0,094 | 0,078 | 0,075 | 0,071 | 0,054 | 0,061 |
| 2003 | 0,088 | 0,094 | 0,069 | 0,093 | 0,081 | 0,072 | 0,100 | 0,088 | 0,078 | 0,075 | 0,062 | 0,057 |
| 2004 | 0,085 | 0,092 | 0,067 | 0,092 | 0,078 | 0,070 | 0,099 | 0,085 | 0,076 | 0,074 | 0,059 | 0,057 |
| 2005 | 0,085 | 0,092 | 0,067 | 0,089 | 0,080 | 0,074 | 0,096 | 0,087 | 0,080 | 0,072 | 0,061 | 0,058 |
| 2006 | 0,084 | 0,094 | 0,066 | 0,087 | 0,082 | 0,076 | 0,097 | 0,091 | 0,086 | 0,069 | 0,062 | 0,063 |
| Average | 0,086 | 0,093 | 0,067 | 0,092 | 0,079 | 0,074 | 0,099 | 0,086 | 0,079 | 0,072 | 0,060 | 0,060 |
| Change | -0,007 | -0,002 | 0 | -0,011 | -0,001 | 0 | -0,005 | 0 | 0,003 | -0,002 | 0,001 | 0 |

Note: (VSEs) less than 20 employees, (SMEs) between 20 and 249 employees, (LEs) between 250 and 4999 employees.

### 4.2 Correlation between the variables

The correlation matrix for the variables is reported in Table (3). The results show that debt is negatively correlated with profitability, but this negative effect is negligible. Moreover, tangibility has also a negative relationship with all profitability ratios. On the other hand, growth opportunities and tax have a positive correlation with profitability. Looking at the relationship between the independent variables themselves, the results show that the multicollinearity is not a problem for the application of analytical techniques[1].

**Table (3)** PEARSON CORRELATION MATRIX

| | PROF1 | PROF2 | ROA | DT | TANG | TAX | VIF |
|---|---|---|---|---|---|---|---|
| PROF2 | 0.909*** | 1 | | | | | |
| ROA | 0.969*** | 0.852*** | 1 | | | | |
| DT | -0.129* | -0.173*** | -0.111*** | 1 | | | 1,04 |
| TANG | -0.102*** | -0.108*** | -0.091*** | 0.124*** | 1 | | 1,03 |
| TAX | 0.401*** | 0.369*** | 0.299*** | -0.166*** | -0.221*** | 1 | 1,02 |
| GROWTH | 0.128*** | 0.109*** | 0.128*** | 0.042*** | -0.013 | 0.059*** | 1,01 |

Note: ***, ** and * represent statistical significance at the 1%, 5%, and 10% levels, respectively.

---

[1] According to test of VIF (Variance Inflation Factor).

### 4.3 Econometric analysis

We estimated the effect of debt on profitability for 2240 French service companies over the period between 1999 and 2006, by using various representatives of profitability ratio as (PROF1, PROF2 and ROA). Moreover, we used two different models (linear and nonlinear) to verify the presence of a nonlinearity of this impact. Furthermore, the estimation was detailed by studying specifically the behavior of companies according to their size (VSEs, SMEs and LEs). So, the results of the estimation of GMM method on panel data models with each of the profitability measures are displayed in Tables (4) to (8).

We can say that all our results are robust for the following reasons: First, the instruments used in our regressions are valid, because Hansen test does not reject the hypothesis of validity of lagged variables in levels and in difference as instruments. Secondly, we note that there is no second-order autocorrelation of errors for difference equation, because the test of second order autocorrelation (AR2) does not allow rejecting the hypothesis of absence of second-order autocorrelation.

We note that debt has no influence on profitability, either in a linear way, or in a non-linear way. In addition, when we present the analysis by using different size classes, we also find that there is no impact of debt on profitability regardless the size of the enterprise. This finding is consistent with Baum et al. (2007) on American industrial companies.

Regarding the control variables, we note, first, that the tangibility negatively affects profitability, this means that companies invest too much in the fixed assets which do not improve their performances, or they do not use their fixed assets efficiently. On the other hand, growth opportunities and tax affect positively the profitability; it means that companies have high profitability level when they have increased growth opportunities and taxes.

**Table (4)** THE EFFECT OF DEBT ON PROFITABILITY *(GMM : TWO STEPS)*

|  | PROF1 | | PROF2 | | ROA | |
|---|---|---|---|---|---|---|
| DT | -0,142 | 0,013 | -0,093 | -0,052 | -0,114 | 0,031 |
|  | (-1,36) | (0,10) | (-0,98) | (-0,43) | (-1,27) | (0,29) |
| DT*2 |  | -0,039 |  | 0,024 |  | -0,043 |
|  |  | (-0,34) |  | (0,21) |  | (-0,42) |
| TANG | -0,020** | -0,018** | -0,022** | -0,021*** | -0,019** | -0,016*** |
|  | (-2,11) | (-2,57) | (-2,53) | (-3,13) | (-2,35) | (-2,87) |
| TAX | 0,149** | 0,205*** | 0,154*** | 0,187*** | 0,078 | 0,127*** |
|  | (2,45) | (10,30) | (2,79) | (9,49) | (1,51) | (7,52) |
| GROWTH | 0,050*** | 0,049*** | 0,041*** | 0,042*** | 0,043*** | 0,042*** |
|  | (6,13) | (6,50) | (5,81) | (6,13) | (6,03) | (6,32) |
| Iyear _2000 | 0,229 | 0,061 | 0,141 | 0,048 | 0,184 | 0,040 |
|  | (1,22) | (0,96) | (0,83) | (0,75) | (1,14) | (0,73) |
| Iyear _2001 | 0,016* | 0,006* | 0,009 | 0,003 | 0,013 | 0,005* |
|  | (1,66) | (1,93) | (1,05) | (1,00) | (1,56) | (1,65) |
| Iyear _2002 | 0,008 | 0,000 | 0,004 | -0,001 | 0,007 | 0,000 |
|  | (1,12) | (0,01) | (0,62) | (-0,51) | (1,07) | (-0,04) |
| Iyear _2003 | 0,010* | 0,003 | 0,005 | -0,001 | 0,008* | 0,003 |
|  | (1,67) | (1,04) | (0,84) | (-0,21) | (1,68) | (1,19) |
| Iyear _2004 | 0,005 | 0,000 | 0,001 | -0,002 | 0,004 | 0,001 |
|  | (1,26) | (0,24) | (0,32) | (-0,80) | (1,37) | (0,49) |
| Iyear _2005 | 0,003 | 0,001 | 0,000 | -0,001 | 0,003 | 0,002 |
|  | (1,08) | (0,45) | (0,09) | (-0,56) | (1,41) | (1,03) |
| Constant | 0,117** | 0,047 | 0,106** | 0,076** | 0,100** | 0,036 |
|  | (2,10) | (1,58) | (2,08) | (2,53) | (2,08) | (1,42) |
| Observations | 10653 | 10653 | 10653 | 10653 | 10653 | 10653 |
| Number of firm | 2240 | 2240 | 2240 | 2240 | 2240 | 2240 |
| Sargan statistic | 2,74 | 14,71 | 4,68 | 23,95 | 3,41 | 14,85 |
| p-value sargan statistic | 0,91 | 0,79 | 0,70 | 0,25 | 0,85 | 0,79 |
| Arellano-Bond test for AR(1) | -9,09 | -12,67 | -9,38 | -12,81 | -9,64 | -12,79 |
| P-value AR(1) | 0,00 | 0,00 | 0,00 | 0,00 | 0,00 | 0,00 |
| Arellano-Bond test for AR(2) | -1,13 | -1,36 | -1,80 | -1,90 | -0,62 | -0,69 |
| P-value AR(2) | 0,26 | 0,17 | 0,07 | 0,06 | 0,53 | 0,49 |

Notes: ***, ** and * represent statistical significance at the 1%, 5%, and 10% levels, respectively. T-students are provided in parentheses. Sargan statistic is a Sargan-Hansen test of overidentifying restrictions. AR (k) is the test for k-th order autocorrelation. Estimation by two-step System GMM. Instruments: (DT) delayed t-3 and t 4, (DT * 2) delayed t-3 and t-4, the rest of explanatory variables are exogenous.

**Table (5)** THE EFFECT OF DEBT ON PROFITABILITY *(GMM : ONE STEP)*

|  | PROF1 | | PROF2 | | ROA | |
|---|---|---|---|---|---|---|
| DT | -0,163 | 0,036 | -0,129 | -0,014 | -0,132 | 0,048 |
|  | (-1,33) | (0,29) | (-1,11) | (-0,11) | (-1,24) | (0,44) |
| DT*2 |  | -0,060 |  | -0,011 |  | -0,063 |
|  |  | (-0,51) |  | (-0,09) |  | (-0,61) |
| TANG | -0,022** | -0,016** | -0,025*** | -0,021*** | -0,021** | -0,016*** |
|  | (-2,14) | (-2,36) | (-2,58) | (-3,05) | (-2,35) | (-2,72) |
| TAX | 0,133* | 0,206*** | 0,130* | 0,186*** | 0,065 | 0,125*** |
|  | (1,85) | (9,98) | (1,90) | (9,19) | (1,04) | (7,05) |
| GROWTH | 0,050*** | 0,048*** | 0,041*** | 0,039*** | 0,043*** | 0,041*** |
|  | (5,96) | (6,42) | (5,71) | (6,05) | (5,90) | (6,33) |
| Iyear _2000 | 0,279 | 0,057 | 0,220 | 0,047 | 0,227 | 0,044 |
|  | (1,24) | (0,85) | (1,03) | (0,71) | (1,16) | (0,76) |
| Iyear _2001 | 0,019 | 0,007** | 0,014 | 0,004 | 0,016 | 0,006* |
|  | (1,63) | (2,02) | (1,23) | (1,16) | (1,53) | (1,89) |
| Iyear _2002 | 0,010 | 0,001 | 0,007 | -0,001 | 0,008 | 0,001 |
|  | (1,16) | (0,25) | (0,87) | (-0,34) | (1,11) | (0,27) |
| Iyear _2003 | 0,011* | 0,003 | 0,007 | 0,000 | 0,010* | 0,003 |
|  | (1,67) | (1,32) | (1,05) | (-0,12) | (1,67) | (1,54) |
| Iyear _2004 | 0,006 | 0,001 | 0,003 | -0,001 | 0,006 | 0,002 |
|  | (1,39) | (0,62) | (0,72) | (-0,51) | (1,49) | (0,95) |
| Iyear _2005 | 0,003 | 0,001 | 0,001 | -0,001 | 0,003 | 0,002 |
|  | (1,15) | (0,53) | (0,36) | (-0,56) | (1,45) | (1,12) |
| Constant | 0,128** | 0,042 | 0,124** | 0,067** | 0,110* | 0,034 |
|  | (1,97) | (1,39) | (2,01) | (2,21) | (1,92) | (1,32) |
| Observations | 10653 | 10653 | 10653 | 10653 | 10653 | 10653 |
| Number of firm | 2240 | 2240 | 2240 | 2240 | 2240 | 2240 |
| Sargan statistic | 2,74 | 14,71 | 4,68 | 23,95 | 3,41 | 14,85 |
| p-value sargan statistic | 0,91 | 0,79 | 0,70 | 0,25 | 0,85 | 0,79 |
| Arellano-Bond test for AR(1) | -8,37 | -12,72 | -8,34 | -12,85 | -8,90 | -12,80 |
| P-value AR(1) | 0,00 | 0,00 | 0,00 | 0,00 | 0,00 | 0,00 |
| Arellano-Bond test for AR(2) | -1,07 | -1,34 | -1,62 | -1,82 | -0,65 | -0,66 |
| P-value AR(2) | 0,29 | 0,18 | 0,10 | 0,07 | 0,52 | 0,51 |

Notes: ***, ** and * represent statistical significance at the 1%, 5%, and 10% levels, respectively. T-students are provided in parentheses. Sargan statistic is a Sargan-Hansen test of overidentifying restrictions. AR (k) is the test for k-th order autocorrelation. Estimation by one-step System GMM. Instruments: (DT) delayed t-3 and t 4, (DT * 2) delayed t-3 and t-4, the rest of explanatory variables are exogenous.

**Table (6)** THE EFFECT OF DEBT ON PROFITABILITY BY SIZE CLASS  *(DEPENDENTE VARIABLE: PROF1)*

|  | (PROF1) |  | TPE |  | PME |  | ETI |  |
|---|---|---|---|---|---|---|---|---|
| DT | -0,142 | 0,013 | -0,186 | 0,091 | -0,126 | 0,044 | 0,008 | -0,014 |
|  | (-1,36) | (0,10) | (-1,49) | (0,61) | (-1,21) | (0,20) | (0,07) | (-0,02) |
| DT*2 |  | -0,039 |  | -0,117 |  | -0,065 |  | -0,074 |
|  |  | (-0,34) |  | (-0,82) |  | (-0,34) |  | (-0,13) |
| TANG | -0,020** | -0,018** | -0,020 | -0,023** | -0,017 | -0,010 | 0,009 | 0,008 |
|  | (-2,11) | (-2,57) | (-1,40) | (-2,04) | (-1,26) | (-1,20) | (0,43) | (0,35) |
| TAX | 0,149** | 0,205*** | 0,068 | 0,165*** | 0,195*** | 0,231*** | 0,238*** | 0,197*** |
|  | (2,45) | (10,30) | (0,73) | (4,49) | (4,46) | (12,51) | (5,40) | (6,07) |
| GROWTH | 0,050*** | 0,049*** | 0,064*** | 0,062*** | 0,032*** | 0,033*** | 0,007 | 0,021 |
|  | (6,13) | (6,50) | (4,01) | (4,17) | (3,00) | (4,77) | (0,37) | (1,12) |
| Iyear_2000 | 0,229 | 0,061 | 0,362* | 0,149* | 0,276 | 0,115 | -0,352* | -0,094 |
|  | (1,22) | (0,96) | (1,70) | (1,77) | (1,22) | (1,27) | (-1,76) | (-0,67) |
| Iyear _2001 | 0,016* | 0,006* | 0,033* | 0,013** | 0,004 | 0,000 | 0,021** | 0,011 |
|  | (1,66) | (1,93) | (1,90) | (2,37) | (0,74) | (-0,02) | (2,00) | (0,88) |
| Iyear _2002 | 0,008 | 0,000 | 0,024* | 0,009** | -0,006 | -0,010*** | 0,012 | 0,011 |
|  | (1,12) | (0,01) | (1,95) | (1,92) | (-1,27) | (-2,85) | (1,47) | (1,14) |
| Iyear _2003 | 0,010* | 0,003 | 0,021** | 0,008** | -0,001 | -0,004 | 0,005 | 0,001 |
|  | (1,67) | (1,04) | (2,23) | (1,99) | (-0,21) | (-1,46) | (0,75) | (0,18) |
| Iyear _2004 | 0,005 | 0,000 | 0,014** | 0,006** | -0,003 | -0,005** | 0,000 | -0,005 |
|  | (1,26) | (0,24) | (2,19) | (1,81) | (-0,92) | (-2,06) | (0,02) | (-0,75) |
| Iyear _2005 | 0,003 | 0,001 | 0,006 | 0,003 | 0,000 | -0,001 | 0,001 | -0,004 |
|  | (1,08) | (0,45) | (1,62) | (1,27) | (-0,12) | (-0,67) | (0,12) | (-0,83) |
| Constant | 0,117** | 0,047 | 0,146** | 0,037 | 0,092** | 0,023 | 0,037 | 0,071 |
|  | (2,10) | (1,58) | (2,19) | (1,02) | (1,70) | (0,39) | (0,55) | (0,30) |
| Observations | 10653 | 10653 | 5760 | 5760 | 4520 | 4520 | 373 | 373 |
| Number of firm | 2240 | 2240 | 1241 | 1241 | 921 | 921 | 78 | 78 |
| Sargan statistic | 2,74 | 14,71 | 2,02 | 22,95 | 4,72 | 11,68 | 4,83 | 20,93 |
| p-value sargan statistic | 0,91 | 0,79 | 0,96 | 0,29 | 0,69 | 0,93 | 0,68 | 0,401 |
| Arellano-Bond test for AR(1) | -9,09 | -12,67 | -6,95 | -9,88 | -6,24 | -8,40 | -2,88 | -1,61 |
| P-value AR(1) | 0,00 | 0,00 | 0,00 | 0,00 | 0,00 | 0,00 | 0,00 | 0,11 |
| Arellano-Bond test for AR(2) | -1,13 | -1,36 | 0,10 | 0,09 | -3,17 | -3,28 | -1,36 | -0,50 |
| P-value AR(2) | 0,26 | 0,17 | 0,92 | 0,93 | 0,00 | 0,001 | 0,17 | 0,62 |

Notes: ***, ** and * represent statistical significance at the 1%, 5%, and 10% levels, respectively. T-students are provided in parentheses. Sargan statistic is a Sargan-Hansen test of overidentifying restrictions. AR (k) is the test for k-th order autocorrelation. Estimation by two-step System GMM. Instruments: (DT) delayed t-3 and t 4, (DT * 2) delayed t-3 and t-4, the rest of explanatory variables are exogenous.

**Table (7)** THE EFFECT OF DEBT ON PROFITABILITY BY SIZE CLASS *(DEPENDENTE VARIABLE: PROF2)*

|  | (PROF2) | | TPE | | PME | | ETI | |
|---|---|---|---|---|---|---|---|---|
| DT | -0,093 | -0,052 | -0,123 | 0,023 | -0,089 | -0,113 | -0,167 | -0,414 |
|  | (-0,98) | (-0,43) | (-1,12) | (0,16) | (-0,93) | (-0,50) | (-1,22) | (-0,62) |
| DT*2 |  | 0,024 |  | -0,036 |  | 0,057 |  | 0,185 |
|  |  | (0,21) |  | (-0,26) |  | (0,29) |  | (0,35) |
| TANG | -0,022** | -0,021*** | -0,026** | -0,029*** | -0,016 | -0,013 | 0,012 | 0,012 |
|  | (-2,53) | (-3,13) | (-2,03) | (-2,62) | (-1,30) | (-1,57) | (0,56) | (0,60) |
| TAX | 0,154*** | 0,187*** | 0,100 | 0,165*** | 0,183*** | 0,199*** | 0,129*** | 0,110*** |
|  | (2,79) | (9,49) | (1,20) | (4,64) | (4,39) | (11,09) | (3,07) | (3,31) |
| GROWTH | 0,041*** | 0,042*** | 0,055*** | 0,056*** | 0,025*** | 0,026*** | 0,025 | 0,023 |
|  | (5,81) | (6,13) | (3,83) | (3,97) | (2,69) | (4,23) | (0,92) | (1,21) |
| Iyear _2000 | 0,141 | 0,048 | 0,263 | 0,121 | 0,158 | 0,083 | -0,197 | 0,020 |
|  | (0,83) | (0,75) | (1,41) | (1,48) | (0,72) | (0,86) | (-0,82) | (0,14) |
| Iyear _2001 | 0,009 | 0,003 | 0,024 | 0,010* | -0,002 | -0,004 | 0,003 | 0,004 |
|  | (1,05) | (1,00) | (1,55) | (1,71) | (-0,32) | (-1,08) | (0,28) | (0,38) |
| Iyear _2002 | 0,004 | -0,001 | 0,017 | 0,006 | -0,009* | -0,011*** | 0,008 | 0,005 |
|  | (0,62) | (-0,51) | (1,58) | (1,21) | (-1,90) | (-3,42) | (0,92) | (0,65) |
| Iyear _2003 | 0,005 | -0,001 | 0,014 | 0,004 | -0,004 | -0,007** | 0,003 | 0,001 |
|  | (0,84) | (-0,21) | (1,58) | (0,97) | (-1,07) | (-2,38) | (0,31) | (0,05) |
| Iyear _2004 | 0,001 | -0,002 | 0,009 | 0,003 | -0,005 | -0,006*** | -0,009 | -0,008 |
|  | (0,32) | (-0,80) | (1,54) | (0,90) | (-1,64) | (-2,70) | (-1,36) | (-1,23) |
| Iyear _2005 | 0,000 | -0,001 | 0,004 | 0,001 | -0,002 | -0,002 | -0,008* | -0,012*** |
|  | (0,09) | (-0,56) | (0,95) | (0,37) | (-0,85) | (-1,19) | (-1,67) | (-2,33) |
| Constant | 0,106** | 0,076** | 0,125** | 0,059* | 0,092* | 0,087 | 0,166* | 0,230 |
|  | (2,08) | (2,53) | (2,13) | (1,66) | (1,86) | (1,45) | (1,95) | (1,14) |
| Observations | 10653 | 10653 | 5760 | 5760 | 4520 | 4520 | 373 | 373 |
| Number of firm | 2240 | 2240 | 1241 | 1241 | 921 | 921 | 78 | 78 |
| Sargan statistic | 4,68 | 23,95 | 4,26 | 28,64 | 5,73 | 13,40 | 4,25 | 14,42 |
| P-value sargan statistic | 0,70 | 0,25 | 0,75 | 0,09 | 0,57 | 0,86 | 0,75 | 0,81 |
| Arellano-Bond test for AR(1) | -9,38 | -12,81 | -7,17 | -9,89 | -7,06 | -8,48 | -2,32 | -2,48 |
| P-value AR(1) | 0,00 | 0,00 | 0,00 | 0,00 | 0,00 | 0,00 | 0,02 | 0,01 |
| Arellano-Bond test for AR(2) | -1,80 | -1,90 | -0,75 | -0,84 | -3,22 | -2,95 | -1,11 | 0,12 |
| P-value AR(2) | 0,07 | 0,06 | 0,46 | 0,40 | 0,00 | 0,00 | 0,27 | 0,90 |

Notes: ***, ** and * represent statistical significance at the 1%, 5%, and 10% levels, respectively. T-students are provided in parentheses. Sargan statistic is a Sargan-Hansen test of overidentifying restrictions. AR (k) is the test for k-th order autocorrelation. Estimation by two-step System GMM. Instruments: (DT) delayed t-3 and t 4, (DT * 2) delayed t-3 and t-4, the rest of explanatory variables are exogenous.

**Table (8)** THE EFFECT OF DEBT ON PROFITABILITY BY SIZE CLASS  *(DEPENDENTE VARIABLE: ROA)*

|  | (ROA) |  | TPE |  | PME |  | ETI |  |
|---|---|---|---|---|---|---|---|---|
| DT | -0,114 | 0,031 | -0,136 | 0,065 | -0,107 | 0,136 | 0,004 | 0,086 |
|  | (-1,27) | (0,29) | (-1,23) | (0,49) | (-1,26) | (0,77) | (0,04) | (0,13) |
| DT*2 |  | -0,043 |  | -0,075 |  | -0,134 |  | -0,149 |
|  |  | (-0,42) |  | (-0,58) |  | (-0,86) |  | (-0,28) |
| TANG | -0,019** | -0,016*** | -0,020* | -0,024** | -0,015 | -0,010 | 0,007 | 0,011 |
|  | (-2,35) | (-2,87) | (-1,74) | (-2,46) | (-1,41) | (-1,53) | (0,37) | (0,52) |
| TAX | 0,078 | 0,127*** | 0,022 | 0,095*** | 0,112*** | 0,144*** | 0,170*** | 0,141*** |
|  | (1,51) | (7,52) | (0,27) | (2,98) | (3,17) | (9,84) | (4,33) | (4,99) |
| GROWTH | 0,043*** | 0,042*** | 0,055*** | 0,053*** | 0,028*** | 0,029*** | 0,002 | 0,015 |
|  | (6,03) | (6,32) | (3,93) | (4,00) | (2,91) | (4,88) | (0,14) | (1,34) |
| Iyear _2000 | 0,184 | 0,040 | 0,273 | 0,116 | 0,242 | 0,105 | -0,287* | -0,084 |
|  | (1,14) | (0,73) | (1,47) | (1,57) | (1,32) | (1,49) | (-1,67) | (-0,62) |
| Iyear _2001 | 0,013 | 0,005* | 0,023 | 0,008* | 0,006 | 0,002 | 0,021** | 0,012 |
|  | (1,56) | (1,65) | (1,54) | (1,68) | (1,14) | (0,51) | (1,96) | (0,91) |
| Iyear _2002 | 0,007 | 0,000 | 0,018* | 0,006 | -0,004 | -0,007** | 0,012 | 0,011 |
|  | (1,07) | (-0,04) | (1,67) | (1,52) | (-1,01) | (-2,52) | (1,52) | (1,21) |
| Iyear _2003 | 0,008* | 0,003 | 0,017** | 0,007* | 0,000 | -0,002 | 0,004 | 0,001 |
|  | (1,68) | (1,19) | (2,02) | (1,88) | (0,12) | (-1,00) | (0,61) | (0,13) |
| Iyear _2004 | 0,004 | 0,001 | 0,012** | 0,006** | -0,002 | -0,004 | 0,001 | -0,004 |
|  | (1,37) | (0,49) | (2,17) | (2,06) | (-0,78) | (-1,90) | (0,26) | (-0,65) |
| Iyear _2005 | 0,003 | 0,002 | 0,006* | 0,004* | 0,001 | 0,000 | -0,002 | -0,004 |
|  | (1,41) | (1,03) | (1,73) | (1,65) | (0,35) | (-0,11) | (-0,44) | (-1,03) |
| Constant | 0,100** | 0,036 | 0,117** | 0,034 | 0,081* | -0,006 | 0,038 | 0,036 |
|  | (2,08) | (1,42) | (1,99) | (1,10) | (1,82) | (-0,13) | (0,63) | (0,18) |
| Observations | 10653 | 10653 | 5760 | 5760 | 4520 | 4520 | 373 | 373 |
| Number of firm | 2240 | 2240 | 1241 | 1241 | 921 | 921 | 78 | 78 |
| Sargan statistic | 3,41 | 14,85 | 1,93 | 21,72 | 4,61 | 10,59 | 4,17 | 22,17 |
| P-value sargan statistic | 0,85 | 0,79 | 0,96 | 0,36 | 0,71 | 0,96 | 0,76 | 0,33 |
| Arellano-Bond test for AR(1) | -9,64 | -12,79 | -7,43 | -10,07 | -6,65 | -8,47 | -2,86 | -1,94 |
| P-value AR(1) | 0,00 | 0,00 | 0,00 | 0,00 | 0,00 | 0,00 | 0,00 | 0,05 |
| Arellano-Bond test for AR(2) | -0,62 | -0,69 | 0,31 | 0,40 | -2,63 | -2,67 | -1,37 | -0,48 |
| P-value AR(2) | 0,53 | 0,49 | 0,75 | 0,69 | 0,01 | 0,01 | 0,17 | 0,63 |

Notes: ***, ** and * represent statistical significance at the 1%, 5%, and 10% levels, respectively. T-students are provided in parentheses. Sargan statistic is a Sargan-Hansen test of overidentifying restrictions. AR (k) is the test for k-th order autocorrelation. Estimation by two-step System GMM. Instruments: (DT) delayed t-3 and t 4, (DT * 2) delayed t-3 and t-4, the rest of explanatory variables are exogenous.

## 5. Conclusion

In this paper, we are interested in the effect of debt on profitability of French service companies. In other words, this article expands the empirical literature regarding the influence of debt on profitability.

There are three essential theories which highlight the influence of debt on corporate profitability, namely: signaling theory, tax theory and the agency costs theory. Furthermore, the disagreement between researchers observed not only theoretically but also empirically.

Lack of studies on French firms and the concentration of studies on listed companies and industrial companies have motivated our study. To do this, we examined empirically the impact of debt on profitability by using the generalized method of moments (GMM) on an unbalanced panel of 2240 French companies of service sector observed over the period 1999-2006. Our sample is composed of unlisted companies like Limited Companies and Limited Liability Companies. In addition, in order to improve the precision of the estimation by reducing heterogeneousness between different sizes of companies, we studied the behavior of these firms according to their size (VSEs, SMEs and LEs). Moreover, we analyzed not only the linear effect of debt on profitability, but also the non-linear effect by estimating a quadratic model which takes into account the squared of debt variable in the regression equation.

According to this study, we can underline that debt has no influence on profitability either in a linear way, or in a non-linear way. This finding is consistent with that of Baum et al. (2007) on American industrial companies. In addition, when we present the analysis using different size classes, we also find that there is no impact regardless the size of enterprise.

For potential research, it would be interesting to take into account some reflections. First, it will be interesting to extend this analysis across different components of corporate debt (long-term and short-term); because, according to most of the studies, contradictory effects have been found. Secondly, we ideally would add new specific variables for companies and sectors, for example, the ownership structure of the corporate capital and the environment in which companies operate. Finally, considering the fact that the relationship between debt and profitability can be non-linear, we can deepen our analysis by using econometric methods that can evaluate the effects of non-linearity as quantile regression and threshold models.